\documentclass[conference]{IEEEtran}
\IEEEoverridecommandlockouts
\usepackage{cite}
\usepackage{amsmath,amssymb,amsfonts}
\usepackage{algorithmic}
\usepackage{graphicx}
\usepackage{textcomp}
\usepackage{xcolor}
\def\BibTeX{{\rm B\kern-.05em{\sc i\kern-.025em b}\kern-.08em
    T\kern-.1667em\lower.7ex\hbox{E}\kern-.125emX}}
\begin{document}

\title{Building Trust in AI-Driven Decision Making for Cyber-Physical Systems (CPS): A Comprehensive Review\\
\thanks{This research was supported by the Science Foundation Ireland and the Department of Agriculture, Food and Marine on behalf of the Government of Ireland Vistamilk research center under grant 16/RC/3835 and by part by Science Foundation Ireland (SFI), co-funded under the European Regional Development Fund under Grant Number 13/RC/2077-P2.}
}

\author{\IEEEauthorblockN{1\textsuperscript{st} Rahul Umesh Mhapsekar}
\IEEEauthorblockA{\textit{Walton Institute, (PAS)} \\
\textit{South East Technological University}\\
Waterford, Ireland \\
rahul.mhapsekar@waltoninstitute.ie}
\and
\IEEEauthorblockN{2\textsuperscript{nd} Muhammad Iftikhar Umrani}
\IEEEauthorblockA{\textit{Walton Institute, (PAS)} \\
\textit{South East Technological University}\\
Waterford, Ireland \\
Muhammad-Iftikhar.Umrani@waltoninstitute.ie}
\and
\IEEEauthorblockN{3\textsuperscript{rd} Malik Faizan}
\IEEEauthorblockA{\textit{Walton Institute, (PAS)} \\
\textit{South East Technological University}\\
Waterford, Ireland \\
faizan.malik@waltoninstitute.ie}
\and
\IEEEauthorblockN{4\textsuperscript{th} Omer Ali}
\IEEEauthorblockA{\textit{Walton Institute, (ENL)} \\
\textit{South East Technology University (SETU)}\\
Waterford, Ireland \\
omer.ali@waltoninstitute.ie}
\and
\IEEEauthorblockN{5\textsuperscript{th} Lizy Abraham}
\IEEEauthorblockA{\textit{Walton Institute, (ENL)} \\
\textit{South East Technology University (SETU)}\\
Waterford, Ireland \\
lizy.abraham@waltoninstitute.ie}

}

\maketitle

\begin{abstract}
Recent advancements in technology have led to the emergence of Cyber-Physical Systems (CPS), which seamlessly integrate the cyber and physical domains in various sectors such as agriculture, autonomous systems, and healthcare. This integration presents opportunities for enhanced efficiency and automation through the utilization of artificial intelligence (AI) and machine learning (ML). However, the complexity of CPS brings forth challenges related to transparency, bias, and trust in AI-enabled decision-making processes. This research explores the significance of AI and ML in enabling CPS in these domains and addresses the challenges associated with interpreting and trusting AI systems within CPS. Specifically, the role of explainable AI (XAI) in enhancing trustworthiness and reliability in AI-enabled decision-making processes is discussed. Key challenges such as transparency, security, and privacy are identified, along with the necessity of building trust through transparency, accountability, and ethical considerations.
\end{abstract}


\section{Introduction}
Advancements in various fields such as agriculture, autonomous systems, healthcare, and transportation have made the interaction between the cyber and physical domains more complex. This complexity has given rise to Cyber-Physical Systems (CPS), which efficiently integrate the cyber and physical domains. CPS entails a closely interlinked communication network where embedded computing devices, smart controllers, physical environments, and humans interact systematically \cite{Intro1}. Artificial Intelligence (AI) and Machine Learning (ML) are crucial in enabling systems to quickly react and make real-time decisions based on new information. AI and algorithmic decision-making have become ubiquitous in modern life, greatly enhancing efficiency and complexity through access to extensive data, sophisticated algorithms, and powerful computing capabilities. However, interpreting these systems has become increasingly challenging due to their complex logic, potentially leading to difficulty in assessing their reliability. Moreover, their limitations, biases, and ethical concerns can render them fragile and unfair \cite{Intro3}.

Agriculture, being one of the oldest and most vital industries globally, faces significant challenges due to the rapid growth of the global population. This surge in population demands more food production and job opportunities. However, traditional farming methods are proving inadequate to meet these escalating needs. Consequently, there is a push towards adopting automated techniques to bridge this gap. Emerging technologies like the Internet of Things (IoT), CPS, Big Data analytics, AI, and ML are revolutionizing agriculture \cite{Intro4}. 

Similarly, recent technological advances have increased the demand for autonomous systems such as road vehicles, Unmanned Aerial Vehicles (UAVs), and robots across various civilian and military applications. Traditionally, operations of these systems are controlled by a human operator, also called human autonomy, where decisions are made by humans\cite{IntroAS1}. On the other hand, integration of AI into these systems not only increases the level of autonomy but also enables them to become more situational aware and make independent decisions, thereby reducing human intervention in task execution \cite{IntroAS2}. Moreover, the decision-making capabilities of these systems, such as prediction and classification, heavily depend on the quality of environmental data acquired through sensors. Therefore, intelligent algorithms can help improve the perception capabilities of these systems and compensate for uncertainties. Recently, various research efforts have been considered to develop resilient and reliable autonomous systems that can be deployed in complex and dynamic environments \cite{IntroAS1}. 

AI has significant potential in healthcare, which incorporates enhancing health through prevention, diagnosis, and management of disease, injury, and various physiological and mental health issues. Medical professionals and associated healthcare sectors govern public health services across primary, secondary, and tertiary care levels \cite{HealthIntro1}. Accessibility to healthcare services is determined by varying social and economic factors as well as differing healthcare policies across nations, leading to challenges due to economic constraints, workforce shortages, and logistical issues. These challenges can compromise the effectiveness of disease diagnosis and subsequent treatment, thus impacting overall health and increasing the risk of mortality. To mitigate these challenges, the adoption of technologies such as sensor fusion in conjunction with AI, ML and data analytics is proposed. Such integrations aim to facilitate more accessible healthcare services and support in the prediction, prognosis, and diagnosis of diseases, ultimately contributing to enhanced healthcare quality and patient well-being \cite{HealthIntro2}.

Integration of AI is necessary to achieve sustainable and efficient CPS, making it crucial to develop more robust and reliant algorithms. This requires seamless integration with classical systems for improved real-time analytics. However, AI systems are sensitive to data, bias and often opaque in their decision-making capabilities, thus termed as "black-box" systems. Some of these factors directly challenge trust, governance, and associated risk factors in CPS domains, thus further hindering their adoption \cite{PdM}. Explainable AI (XAI), an emerging subset of AI, accounts for these trustworthiness aspects of AI. This article provides several use-cases where the integration of XAI in CPS domains may provide enhanced explainability of AI actions to increase trust \cite{XAItrust, trustXAI}.

\section{Related Works}

This review aims to develop trust in AI-enabled decisions for CPS across various use cases, such as Agriculture, Autonomous Systems, and Healthcare. Some reviews discuss advancements in AI for CPS and the associated challenges. Babajide \textit{et al.} in \cite{RW1} discussed that the integration of AI and ML with IoT and CPS is essential for their advancement, ensuring efficient operations and dependable systems. Challenges such as data fusion, security, and privacy persist, requiring innovative solutions such as blockchain, and edge computing. Data fusion challenges include handling diverse modalities, proper calibration for accurate fusion, addressing trivial data, and synchronizing operational timing for real-time applications. Security remains a significant concern due to the vast number of interconnected devices, necessitating robust AI and ML systems to counter adversarial attacks and protect consumer data. Similarly, Jiyeong \textit{et al.} \cite{RW2} has discussed the integration of AI with Industrial Cyber-Physical Systems (ICPS) for optimization of the process. It proposes the concept of AI-augmented ICPS (AICPS) and discusses its components, interactions, and design considerations. The advancements in industrial AI technologies are examined for their applications in AICPS, highlighting benefits such as real-time monitoring and management.  Similarly, utilizing Smart Agriculture (SA) as a CPS can increase the efficiency of farms by gathering real-time data on weather conditions, crop health, and soil quality. Various use cases such as irrigation and solar energy systems have benefited from SA integration which uses sensor and actuator networks that enable conservation of energy and maintain optimal conditions for plant growth \cite{RWAG1}. 

From the perspective of a short review briefly discussing the the role of advanced computing technologies for enhancing the capabilities of autonomous systems has been presented by Chen \textit{et al.} in\cite{RWAS1}. This study delves into the state-of-the-art of autonomous systems and explores their integration with federated learning, IoT, and big data. It also discusses the evolution and adaptation of ML algorithms addressing uncertainties, and improving communication and control capabilities using IoT and big data for real-time data processing in autonomous systems. The authors of \cite{RWAS2} investigate the utilization of potential computational technologies in designing trustworthy AI systems. It introduces six important dimensions to design such systems: safety and robustness, fairness, explainability, privacy, and accountability. In addition to that, it examines the key technologies and their applications within each dimension across real world AI systems. Moreover, the work underscores the significance of mitigating bias issues within AI algorithms to ensure fairness in practical AI applications. It also offers a comprehensive overview of various mitigation techniques aimed at counteracting biases in AI systems.

The fusion of technological advancements in Wireless Sensor Networks (WSN), medical sensors, and cloud computing (CC) presents CPS as a significant option for various healthcare applications, including patient services within hospitals and at home \cite{RWH1}. CC enables efficient management and analysis of larger dataset related to the healthcare information collected by CPS, often characterized by large volumes, complex structures, and diverse characteristics. Milenkovic \textit{et al.} in \cite{RWH2} proposed alternative data management solutions utilizing NoSQL databases to improve the processing, storage, indexing, and analysis of healthcare information, mitigating the limitations of traditional relational databases. The integration of smart devices with advanced sensing and network capabilities, such as wearable technology and lab-on-chip meters, contributing towards smart healthcare \cite{RWH3}. In particular, wireless body area networks (WBANs) are recognized as a leading technology that provides seamless healthcare services and facilitates continuous health monitoring. These networks form a subset of medical CPS capable of functioning in complex environments like hospitals, addressing communication challenges in various medical scenarios \cite{RWH2}.
While many studies center on the technological landscape of AI and its integration with CPS in various domains, there is a gap in addressing the trust in AI system, thus served as the primary motivation for this article. This review makes the following contributions:

\begin{itemize}
    \item{A comprehensive account of AI-enabled CPS applications}
    \item {Identification of the need for trust in AI-enabled applications and their decision-making process}
    \item{The need to integrate XAI for establishing trust, transparency and granularity in AI-decision making}
\end{itemize}

\section{AI for CPS domains}
\begin{figure*}[ht]
\centering
\includegraphics[width=\textwidth]{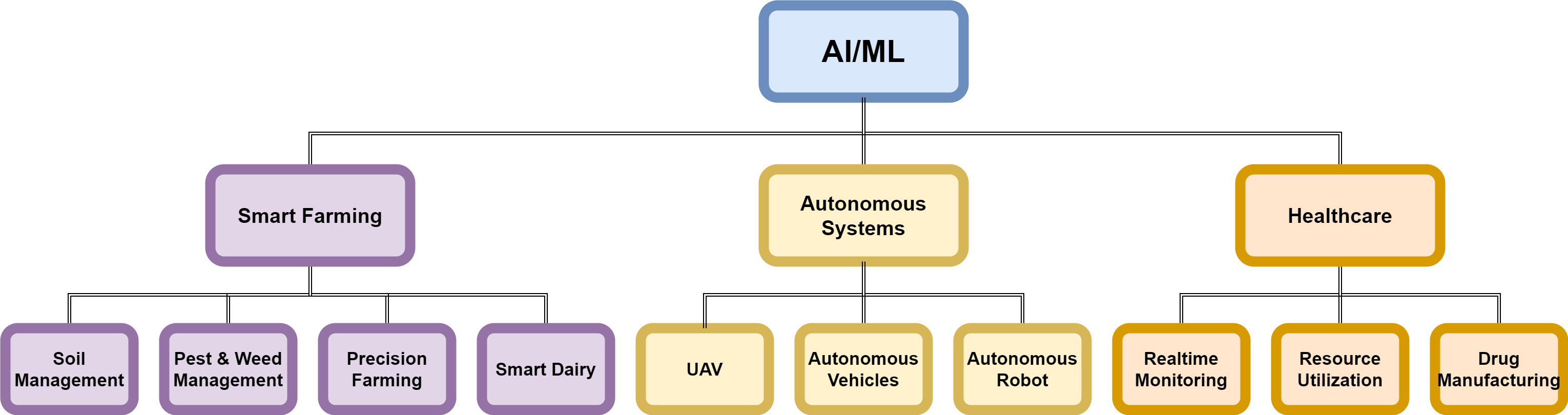}
\caption{Applications of AI in proposed use cases}
\label{Fig.1}
\end{figure*}

This section discusses the applications and benefits of using AI in the use cases considered in this study. Fig. \ref{Fig.1} shows the applications of AI in Smart Farming, Autonomous systems and Healthcare which is the focus of this research.

\subsection*{\textbf{Use of AI and CPS in Agriculture:}}

AI plays a pivotal role in agricultural management across various domains \cite{AgriAI1}. In soil management, it utilizes techniques like management-oriented modeling and artificial neural networks for predictive performance on soil parameters. Similarly, in pest and weed management, AI aids in the accurate monitoring and identification of infected plants through techniques such as digital image analysis and sensor ML. Disease management benefits from AI by enabling early detection through image analysis of crops and animals. Crop management is revolutionized with AI recommendations tailored to crop choices and resource allocation. Water-use optimization is enhanced through AI-managed irrigation systems, integrating real-time data for efficient resource utilization \cite{AgriAI1}. Precision Farming is another application area where CPS and AI can be leveraged to improve the results. It integrates computational intelligence into complex systems, such as precision agriculture, using cooperative sensor swarms and layered architectures. Researchers have proposed CPS frameworks for precision spraying, monitoring, and decision-making in agriculture, using WSN and robots to enhance efficiency and involve human operators for improved performance \cite{AgriAI5}. 

Similarly, it has been used in the dairy industry for several applications such as disease detection in cows (clinical mastitis, ketosis, metritis, etc). CPS has significantly contributed towards smart dairy farming. Farm Management System (FMS) is one of the applications that serves as the key element of dairy farm optimization, integrating vital functions like herd management, milk production monitoring, and financial administration into a centralized software solution. This system not only streamlines operations but also enhances decision-making through data analysis\cite{AgriD2}. Deep Learning (DL) can improve Milk Quality Analysis (MQA) for milk adulteration detection \cite{AgriAI4}. However, there has been a lack of DL implementation in MQA due to lack of trust in the DL predictions. While AI offers numerous advantages in agriculture, addressing security and privacy concerns is crucial to build trust in the use of CPS and AI within the agricultural sector.
\subsection*{\textbf{Role of AI in Autonomous Systems:}}
AI holds significant promise for enhancing the reliability and safety of autonomous systems across various applications, such as smart cities, remote sensing, rescue operations, transportation, and others \cite{AIAS1}. In the domain of UAV-based precision agriculture, AI techniques such as data fusion and DL play a crucial role for improving the overall accuracy of crop yield prediction \cite{AIUAVPA1}. Similarly, DL aids in accurately categorizing the variety of seasonal crops that are cultivated within specific fields \cite{AIUAVPA2}. This information is valuable for various tasks, like agriculture insurance, predicting market trends, land leasing decisions, and supply chain management. Moreover, unique framework has been developed for AI-enabled package delivery drones to avoid failures in package delivery missions \cite{AIUAVPD}. It uses a neural network regression model to monitor the battery consumption of drones in a real-time environment. To minimize infrastructure damage in flood-affected areas, Convolutional Neural Networks (CNNs) algorithm was deployed in UAVs to detect flood-affected zones \cite{AIUAVDM}. In this scenario, the UAVs captured aerial imagery, which was later used to extract various flood-related features, including damaged infrastructure. Additionally, AI techniques can bring intelligence in industrial robots to improve their operational and cognitive capabilities. For example, AI techniques were used to design an autonomous product-quality evaluation system. This system monitors the product quality and distinguishes between defective and non-defective products without human intervention \cite{AIROBMAN1}. Similarly, these techniques have been used to find anomalies in robotic data, predict potential failures, reducing downtime, and maintenance costs \cite{AIROBMAN2}. In smart cities, the inclusion of AI enables the development of intelligent transportation systems. This targeted deployment has the potential to detect real-time traffic, ensuring safety for public and autonomous vehicles \cite{AIAV}. Furthermore, CPS ensures human safety in human-robot collaborative operations \cite{AICPS2}. This system utilizes sensors to continuously monitor the environment and trigger an alarm if the safety distance between humans and robots is below the threshold. Although AI and CPS have promising capabilities to design intelligent autonomous systems, but the unpredictable behavior of AI systems could significantly deteriorate the performance of these systems. 

\subsection*{\textbf{Role of AI and CPS in Healthcare:}}
After the COVID-19 pandemic, a significant increase has been observed in the adoption of AI technologies for healthcare. CPS plays a vital role in healthcare beyond hospitals, extending to assisted living and elderly care. Its application in healthcare is generally categorized into two areas. In \textit{Assisted application}, health monitoring is conducted without limiting the daily activities of an individual. Real-time physiological data is acquired through Photoplethysmography (PPG) or biosensors, providing medical guidance to patients. This opens up opportunities to remotely support and assist many individuals, particularly the elderly \cite{HealthAI1}. In \textit{controlled application}, a rapid and extensive level of support is offered, similar to the facilitates in hospitals and intensive care units. This rapid and extensive level of support involves integrating data from multiple sources, which include body sensors, bedside monitors, and clinical observations. The information data is combined as a closed-loop network with human intervention for decision-making. The combination of these two healthcare application areas transforms the healthcare system into one complex and vast safety-critical CPS. It also offers many benefits like improved physiological status, patient safety, drug research and development, and medical flow \cite{HealthAI2}, \cite{HealthAI3}.

\section{Challenges of using CPS and AI in Agriculture, Autonomous Systems and Healthcare}
\begin{figure*}[ht]
\centering
\includegraphics[width=\textwidth]{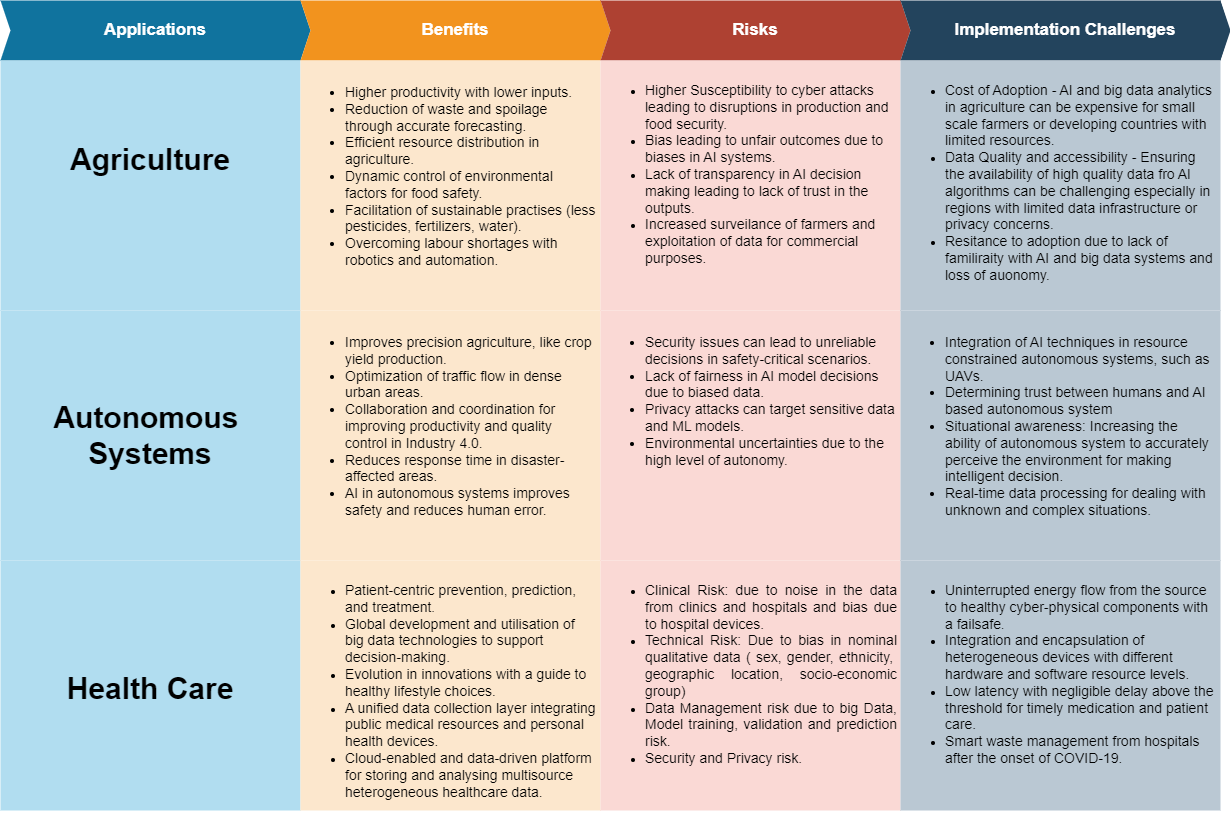}
\caption{Summary of the studies used in the review}
\label{Fig. 2}
\end{figure*}

Several challenges have been identified based on the applications and benefits associated with CPS and AI integration in the above-mentioned use cases. This section explores the challenges in integrating AI and CPS for the applications used in this study. Fig \ref{Fig. 2}. shows the benefits and challenges associated to understand the tradeoffs for using CPS and AI in the specific application areas. 

\subsubsection*{\textbf{Challenges associated with CPS and AI in Agriculture}}
Transparency is crucial in AI, revealing how a system makes decisions. In agriculture, Artificial Neural Networks (ANNs) are common for their adaptability in remote sensing, despite opacity and high costs as complexity rises. Balancing transparency, computational efficiency, and performance is vital in selecting ML techniques for agricultural applications, notably in remote sensing and precision agriculture. AI biases from skewed data or excluding diverse farming practices pose challenges, causing injustices and disadvantages to small-scale farmers. Opacity in ML systems raises trust and responsibility concerns, with farmers facing dilemmas over contradictory AI recommendations \cite{AGRICH2}. Cybersecurity is critical in modern agriculture, especially concerning system security and data privacy. With the rise in interconnected computers and electronic devices in agricultural production facilities, assessing cybersecurity risks is necessary. This involves considering all components and data communication, alongside detailing use cases and stakeholder roles for a comprehensive cybersecurity assessment \cite{AGRICH3}. CPS integration also introduces vulnerabilities to hacking and cyber-attacks, similar to other technological domains. Instances of ransomware attacks in various industries highlight the potential for significant damage, particularly as agricultural infrastructure becomes more electronically reliant. Moreover, the concentration of ownership in the agricultural sector increases susceptibility to cyber threats, with potential ramifications extending to national food security and social order \cite{AGRICH2}. Resolving these issues is vital for using AI in agriculture for benefits, as reluctance may hinder progress.
\\
\subsubsection*{\textbf{Challenges associated with AI in autonomous systems}}

Similarly, extensive research has been carried out over the years to investigate the potential challenges involved in designing, developing, and deploying the robust and reliable autonomous system \cite{IntroAS1, ASCHAL2, ASCHAL3}. These studies have identified both technical and non-technical challenges, such as security issues, sensor noise and bias, sensor fusion techniques, sensor data quality, fault tolerance, real-time data processing, data integration, trust management, and system-human collaboration. Addressing these challenges is imperative for the development of robust, trustworthy, and reliable autonomous systems.

\subsubsection*{\textbf{Challenges associated with CPS for healthcare}}

The challenges associated with CPS in healthcare are complex and critical for optimizing performance and the safety of patient care. Representation of medical device systems in CPS models integrates complex and advanced treatment algorithms that interact with the physical system components, notably the patient\cite{HealthAI2}. Their growing complexity and connectivity, pose security, safety and dependability challenges for CPS-based medical devices. The confidentiality and integrity of patient data are essential to maintain privacy and security in healthcare. Moreover, unauthorized access or misuse of such data can lead to reputational damage, loss of privacy, mental distress, and subsequent health deterioration \cite{HealthAI5}. Similarly, CPS keeps incorporating up-to-date versions of IoT \cite{HealthAI6} by integrating WSN \cite{RWH1}, machine-to-machine (M2M) communication, radio frequency identification, pervasive computing technology, network communication tools, and novel control models \cite{HealthAI3}. These CPS applications benefit significantly from the inclusion of intelligent devices and wireless networks that facilitate sophisticated services informed by data from the physical world \cite{RWH2}. A primary concern for handling large-scale healthcare data involves establishing a robust distributed storage architecture that can process and analyze large volumes of data \cite{HealthAI7}. These limitations can delay complex query processing, reducing data transmission loads and enabling context-sensitive predictive modeling \cite{HealthAI3}. Moreover, there is a demand for high-performance CPS systems which include IoT systems, incorporating immediate system feedback loops and ensuring interoperability amongst medical devices \cite{HealthAI6},\cite{HealthAI7}.

\section{Building Trust in AI and CPS for applications}
\begin{figure*}[ht]
\centering
\includegraphics[width=\textwidth]{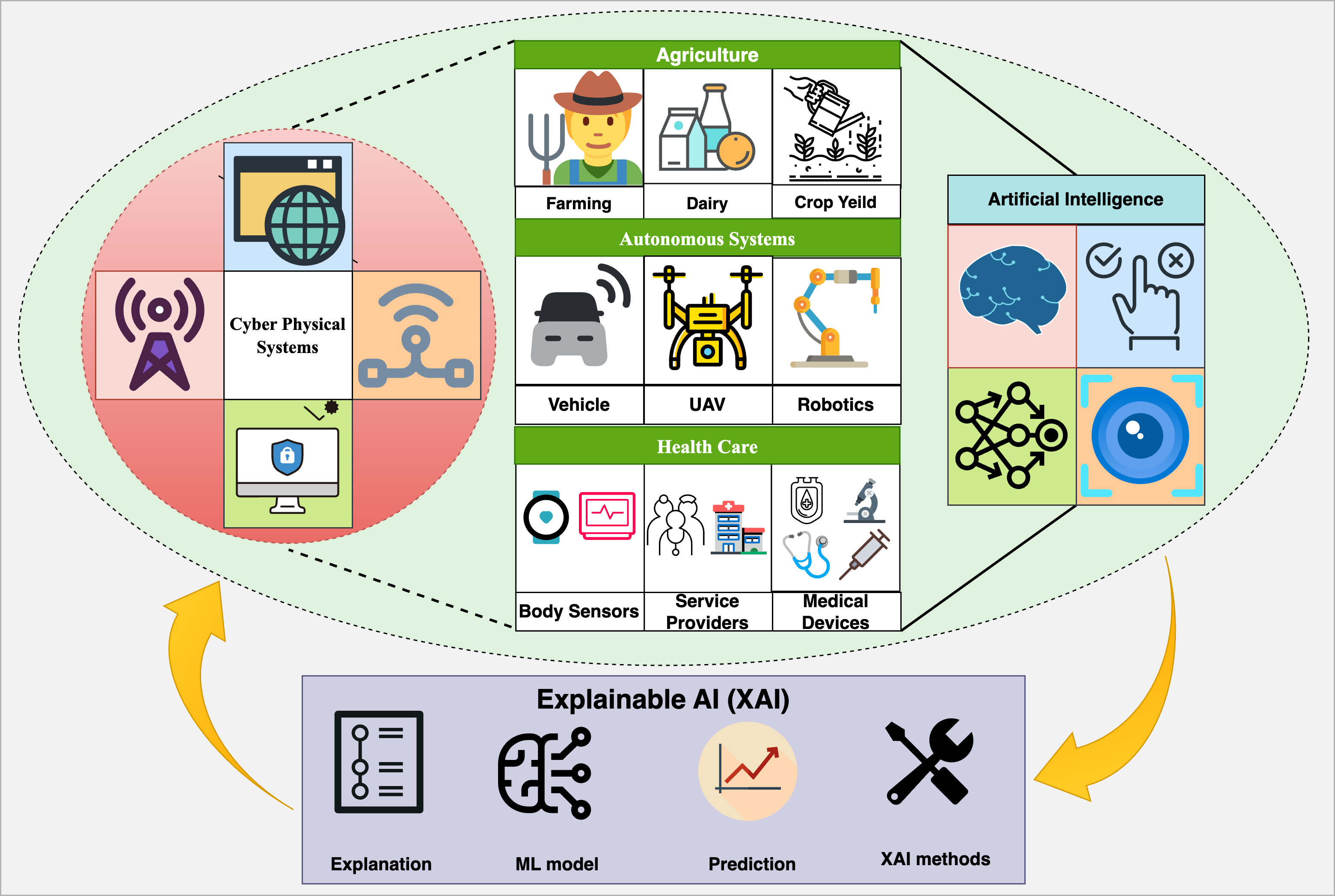}
\caption{The technological landscape of explainable AI integration with CPS.}
\label{Fig.3}
\end{figure*}

Fig. \ref{Fig.3} shows the technological landscape of XAI integration for building trust in AI and CPS applications.

\subsubsection*{\textbf{Building Trust for IoT applications in Agriculture}}
Risks associated with AI in agriculture, such as bias, lack of transparency, and vulnerability to hacking, require attention from designers and policymakers. Solutions to mitigate these risks include ensuring diverse perspectives in AI training data, developing explainable AI, and implementing robust cybersecurity measures. Additionally, governments may encourage local data collection, incorporate indigenous agricultural knowledge, and establish regulations to address privacy, surveillance, and data ownership concerns \cite{AGRICH2}.

A deeper understanding of AI, ethics, and stakeholder interests, along with evidence-based approaches, is crucial for navigating the complexities of AI in agriculture. Despite the potential of intelligent automation in the agriculture domain, the risks of inheriting trust deficiency may overshadow its applicability. First, the lack of data ownership and privacy may deeply affect the farming community as the AI-based models may require insights from a broader region which would eventually deprive the trade originality for farmers. Next, a mere bias in the algorithm may cause additional usage of resources, resulting in additional costs depending on the region. This bias would eventually translate into varied produce pricing thus further overburdening the community. Furthermore, the adoption of AI specifically in rural communities may pose additional challenges; cost, learning curve, and adaptability to name a few. Therefore, a deeper insight into AI decision-making capabilities and the overall impact must be presented to the community for adoption. In addition to that, a more curated use case must be presented for better understanding, thus further advocating the need for XAI in the smart agriculture domain. In regard to the adoption of these technologies, the overwhelming dependency on these systems may marginalize autonomy in the process, thus further diminishing the required trade skills. Finally, the long-term implications, such as biodiversity, use of technology, land usage changes, and continuous integration and dependency must be thoroughly addressed. Therefore, in light of these challenges, the integration of XAI in Smart Agriculture domain may promise to establish the trust and governance required.

\subsubsection*{\textbf{Building Trust in Autonomous systems}}
The integration of AI models enhances the operational capabilities of autonomous systems, but it also brings significant trust challenges. Reliability and robustness of these system depends on data quality, trust in AI models, decision-making processes, and responses to uncertainties \cite{ASTRST}.

The AI-powered autonomous systems calls for high degree of trust. As these systems are becoming increasingly integrated into our modern lives, ranging from medical diagnosis to self-driving cars and even high-investment high-frequency trading; a trust deficit can bring dire consequences. It underpins the end-user as well as confidence in the operation, reliability, and capability of these systems. It is crucial to understand that not the trust factor but the ability to understand the model behavior and its decisions can enable long-term fully integrated trustworthy AI systems. Imagine the absence of trust in every day use-cases around us. For instance, a self-driving vehicle that fails to detect a pedestrian crossing the road; an inaccurate algorithm providing untrustworthy medical diagnosis, and even an AI algorithm that fails to accurately invest in new emerging stock could potentially end up causing billions in damages, and in worst case loss of lives. XAI can provide deeper insights into AI behavior, its feature selection, ability to adapt to future problems, and most importantly the rationale behind its decisions. 

\subsubsection*{\textbf{CPS Trust building for healthcare }}
CPS malfunctions lead to significant economic consequences and disrupt the functionality of associated systems. These systems require physical infrastructure and complex networking models. Because of diverse healthcare applications, CPS are becoming more flexible, reliable, and complex. Enhancement to CPS architecture offers significant improvement in adaptability, scalability, usability, security, and safety. These advancements mark a significant transition from conventional embedded systems. CPS must establish trust in healthcare by providing transparent status and readings, ensuring real-time performance. 

Amongst all other domains, the application of AI in healthcare requires a great degree of trust, privacy, security and strict governance policies. The underlying insights of AI  working are even more critical in this domain as compared to other sectors. Modern healthcare systems heavily rely on the use of AI models ranging the entire spectrum from diagnosis to prognosis including initial diagnosis, treatment planning, insurance cover management, treatment, and patient care. In this regard, trust in these systems cannot merely be treated as a matter of convenience but in fact a matter of life and death. For instance, consider an AI system that can classify a tumor with a very high degree of accuracy but ruling out its malignant or benign nature could potentially result in delay of treatment, and in worst cases the complete absence of any treatment resulting in life-threatening scenarios. Additionally, modern AI systems often advise on patient re-admissions or even the probability of complications during surgery. If the decision-making capabilities are opaque and are not validated, healthcare providers may completely ignore to act upon them thus missing early intervention possibilities. XAI in this regard may play a crucial role for identifying, demystifying and addressing the shortcomings of this "black-box" approach for AI models. With XAI integrated into the AI systems, physicians and clinical aid can better understand how a model reaches a certain conclusion, and whether the high-accuracy figures actually relate to medical knowledge and are free from bias and error. This will ensure collaborative communication between humans and future AI models to gradually build trust and confidence in AI-based decisions to complement rather than supplant human knowledge, expertise, and judgment. 

\section{Conclusion}
The paper explores the integration of Artificial Intelligence in Cyber-Physical Systems, focusing on agriculture, autonomous systems, and healthcare domains. In agriculture, use cases include soil management, pest and weed management, precision farming, and smart dairy. For autonomous systems, use cases cover UAVs, autonomous vehicles, and robotics. In healthcare, use cases involve real-time health monitoring, resource utilization, and drug manufacturing. The study aims to identify challenges such as security and data privacy across these applications and discusses tradeoffs between advantages and implementation challenges. Furthermore, it underscores the importance of cyber security, transparency, fairness, and trustworthiness in AI-based decision-making and reviews the role of Explainable AI (XAI) in addressing these challenges and enhancing trust and reliability in AI-based decisions.

\end{document}